\documentclass{aastex}
\usepackage{emulateapj5}

\newcommand{\Halpha}{\hbox{\rm H$\alpha$}}
\newcommand{\kms}{${\rm km~s}^{-1}$}

\slugcomment{Accepted for publication in the {\em ApJ}, 5 March 2001.}

\shorttitle{Intergalactic Gas in NGC 2442 Group}
\shortauthors{Ryder et al.}

\begin{document}


\title{HIPASS Detection of an Intergalactic Gas Cloud in the NGC~2442
Group}


\author{S. D. Ryder\altaffilmark{1},
B. Koribalski\altaffilmark{2},
L. Staveley-Smith\altaffilmark{2},
V. A. Kilborn\altaffilmark{3},
D. F. Malin\altaffilmark{1},
G. D. Banks\altaffilmark{4},
D. G. Barnes\altaffilmark{5},
R. Bhatal\altaffilmark{6},
W. J. G. de~Blok\altaffilmark{2},
P. J. Boyce\altaffilmark{4},
M. J. Disney\altaffilmark{4},
M. J. Drinkwater\altaffilmark{3},
R. D. Ekers\altaffilmark{2},
K. C. Freeman\altaffilmark{7},
B. K. Gibson\altaffilmark{5},
P. A. Henning\altaffilmark{8},
H. Jerjen\altaffilmark{7},
P. M. Knezek\altaffilmark{9},
M. Marquarding\altaffilmark{3},
R. F. Minchin\altaffilmark{4},
J. R. Mould\altaffilmark{7},
T. Oosterloo\altaffilmark{10},
R. M. Price\altaffilmark{2,8},
M. E. Putman\altaffilmark{7},
E. M. Sadler\altaffilmark{11},
I. Stewart\altaffilmark{2,12},
F. Stootman\altaffilmark{6},
R. L. Webster\altaffilmark{3},
and A. E. Wright\altaffilmark{2}
}
\affil{~~~}
\email{sdr@aaoepp.aao.gov.au}

\altaffiltext{1}{Anglo-Australian Observatory, P.O. Box 296, Epping, NSW~1710,
Australia.}
\altaffiltext{2}{Australia Telescope National Facility, CSIRO, P.O. Box 76,
Epping, NSW~1710, Australia.}
\altaffiltext{3}{University of Melbourne, School of Physics, VIC~3010,
Australia.}
\altaffiltext{4}{University of Wales, Cardiff, Department of Physics \&
Astronomy, P.O. Box 913, Cardiff CF2 3YB, U.K.}
\altaffiltext{5}{Centre for Astrophysics and Supercomputing,
Swinburne University of Technology, P.O. Box 218, Hawthorn, VIC 3122,
Australia.}
\altaffiltext{6}{University of Western Sydney Macarthur, Department of Physics,
P.O. Box 555, Campbelltown, NSW~2560, Australia.}
\altaffiltext{7}{Research School of Astronomy \& Astrophysics, Mount Stromlo
Observatory, Cotter Road, Weston, ACT~2611, Australia.}
\altaffiltext{8}{University of New Mexico, Department of Physics \& Astronomy,
800 Yale Blvd. NE, Albuquerque, NM~87131.}
\altaffiltext{9}{Space Telescope Science Institute, 3700 San Martin Drive,
Baltimore, MD~21218.}
\altaffiltext{10}{ASTRON, P.O. Box 2, 7990 AA Dwingeloo, The Netherlands.}
\altaffiltext{11}{University of Sydney, School of Physics A28, Sydney,
NSW~2006, Australia.}
\altaffiltext{12}{X-ray Astronomy Group, Department of Physics \& Astronomy,
University of Leicester, Leicester LE1 7RH, U.K.}


\begin{abstract}
We report the discovery, from the \ion{H}{1} Parkes All-Sky Survey
(HIPASS), of a gas cloud associated with the asymmetric spiral
galaxy NGC~2442.  This object, designated HIPASS J0731--69, contains
$\sim10^{9}$~M$_{\odot}$ of \ion{H}{1}, or nearly one-third as much
atomic gas as NGC~2442 itself.  No optical counterpart to any part of
HIPASS J0731--69 has yet been identified, consistent with the gas
being diffuse, and with its stream-like kinematics. If the gas in
HIPASS J0731--69 was once part of NGC~2442, then it was most likely a
fairly recent tidal encounter with a moderately massive companion
which tore it loose, although the possibility of ram-pressure
stripping cannot be ruled out. This discovery highlights the potential
of the HIPASS data for yielding new clues to the nature of some of
the best-known galaxies in the local universe.
\end{abstract}




\addtocounter{footnote}{-12}
\section{Introduction}

The nearby southern galaxy NGC~2442 ($V_{\odot}=1449$~\kms) provides a
striking example of well-developed, but asymmetric spiral arms
(Figure~\ref{f:h1ondss}; see also Panel 207 of \citet{sb94}). The
northern arm bends back on itself quite tightly, and is sharply
bisected by one continuous dust lane, while the southern arm is
shorter, more open, and criss-crossed by numerous dust features. There
have been several attempts to ascribe this asymmetry to a past
interaction with a neighboring galaxy, but no clear culprit has yet
emerged. \citet{sb94} suggested the nearby E0 galaxy NGC~2434
($V_{\odot}=1390$~\kms, ${\rm separation}=16\farcm8$) or perhaps the
more distant spiral NGC~2397 ($V_{\odot}=1363$~\kms, ${\rm
separation}=85\arcmin$); \citet{elm91} also blamed NGC~2434, but
\citet{mb97} argued on the basis of its proximity, and slightly
disturbed morphology, that the small SB0/a galaxy AM~0738--692 is a
more likely candidate. In an \ion{H}{1} aperture synthesis study of
NGC~2442 with the Australia Telescope Compact Array (ATCA),
\citet{sjh98} failed to detect any \ion{H}{1} associated with AM
0738-692, but did find evidence of weakly-disturbed \ion{H}{1} in the
irregular galaxy ESO~059-G006 ($V_{\odot}=1346$~\kms) located some
$17\farcm1$ south of NGC~2442.

A deep optical image (Figure~\ref{f:dfm}) showing NGC~2442 and most of
the galaxies mentioned above has been produced by adding three
contrast-enhanced IIIaJ plates from the UK Schmidt Telescope. This
image also highlights a diffuse extension of the northern arm, which
extends almost halfway towards ESO~059-G006 before becoming
indistinguishable from the diffuse Galactic reflection nebulosity
which pervades the field.

Distance estimates in the literature for NGC~2442 range from 14~Mpc
\citep{sd93}, to 17.1~Mpc \citep{nbg88}. We adopt $D=15.5$~Mpc
($1\arcmin=4.5$~kpc) in what follows.  \citet{lgg93} assigns only
three other galaxies to the group (LGG 147) including NGC~2442:
NGC~2397; NGC~2434; and PGC~20690 ($={\rm AM}0720-723$), a Magellanic
irregular some $3\fdg5$ south of NGC~2442. The same four galaxies were
similarly grouped together by \citet{nbg88} and designated
53--17. However, in light of the above discussion, and several recent
redshift determinations, it is clear that the NGC~2442 group is
somewhat more populous than once thought. Table~\ref{t:neigh} lists
all the galaxies within $90\arcmin$ (400~kpc projected distance) of
NGC~2442 whose redshifts are known. The galaxies ESO~059-G010 and
ESO~059-G007 are probably also close neighbors, but no redshifts are
yet available.

In this paper, we report the discovery of significant amounts of
diffuse \ion{H}{1} to the northwest of NGC~2442, and discuss how
this may help shed new light on the nature of its unusual appearance.

\section{Observations and Results\label{s:obs}}

The \ion{H}{1} Parkes All-Sky Survey (HIPASS) is a blind \ion{H}{1}
survey of the entire local southern sky ($\delta < 2 \degr$, $-1200 <
V_{\odot} < 12700$~\kms), carried out between February 1997 and March
2000. The survey uses a 13-beam receiver mounted on the Parkes 64~m
telescope to scan the sky in $8 \degr$ strips of declination,
revisiting each point on five separate occasions, for an effective
integration time of 460~s~beam$^{-1}$. Full details of the multi-beam
receiver system, survey strategy, and on-line data analysis can be
found in \citet{lss96} and \citet{dgb01}. The HIPASS
datacubes\footnote{Spectra from the HIPASS data, gridded at $8\arcmin$
intervals, are now publicly accessible from
\anchor{http://www.atnf.csiro.au/research/multibeam/release/}{http://www.atnf.csiro.au/research/multibeam/release/}.}
used in this analysis have a final beamwidth of $\sim 15\farcm5$,
velocity resolution 18~\kms, r.m.s. noise level of $\sim
13$~mJy~beam$^{-1}$, and a spatial pixel size of $4\arcmin$.

One of the first major data products to emerge from HIPASS is the {\em
HIPASS Bright Galaxy Catalog\/} \citep{bgc01}. In the course of
measuring the centroid, total flux, and extent of \ion{H}{1} emission
from the 1000~brightest galaxies in the HIPASS data, we noticed an
unusual extension to the integrated intensity (zeroth moment of the
\ion{H}{1} datacube) image of NGC~2442, with no obvious optical
counterpart (Figure~\ref{f:h1ondss}). The peak \ion{H}{1} column
density in this region is just 0.1~M$_{\odot}$~pc$^{-2}$, or
$1.3\times10^{19}$~cm$^{-2}$. In accordance with the IAU standard, we
assign this new object the provisional designation HIPASS J0731--69.

From an analysis of the zero moment map, as well as the total
\ion{H}{1} profile (Figure~\ref{f:h1prof}), we find this new source to
have a flux integral of $18\pm3$~Jy~\kms\ (see \citet{dgb01} for a
discussion of the uncertainties in total fluxes from the HIPASS
data). Although the \ion{H}{1} profile is somewhat asymmetric (and
most unlike the ``double-horned'' or Gaussian profile expected of a
spiral or a dwarf galaxy) we derive a heliocentric velocity for the
peak of the line emission of $1481\pm18$~\kms, and profile width at
20~\% of the peak intensity $W_{20}=270\pm60$~\kms. For comparison,
the equivalent HIPASS parameters for NGC~2442 itself, from a robust
moment analysis and from the \ion{H}{1} profile (also slightly
asymmetric -- Fig.~\ref{f:h1prof}), are a flux integral of
$64\pm4$~Jy~\kms, heliocentric velocity of $1456\pm40$~\kms, and
$W_{20}=550\pm12$~\kms.

The similarity in \ion{H}{1} profile shapes raises the concern that
HIPASS J0731--69 may just be a ``ghost'' of NGC~2442. We examined the
five separate scans which pass through this region (each of which is
displaced from the previous scan by $7\arcmin$; Barnes et~al.  2001),
and found a firm detection in each one of them, confirming that HIPASS
J0731--69 is not an artifact of either the scanning, or the gridding
process.

Of the 1000 or so brightest HIPASS sources examined by \citet{bgc01},
NGC~2442 was the only object found to have such a notable extension,
with no optical counterpart. Most other extended sources turn out to
be associated with known loose groups of galaxies.  At least two
sizable \ion{H}{1} clouds lying outside the Zone of Avoidance, but
apparently devoid of any stars, have also been found by visual
inspection of the HIPASS data \citep{kil00,bk01}, but with velocities in
the range $400-500$~\kms, they are probably part of, or just outside
of the Local Group.

In an attempt to further characterize the nature of HIPASS J0731--69,
we used the ATCA in its 750A configuration on 2000~August~25 UT to map
the field of this emission. A mosaic of four pointings was observed
over 10~hours, extending $\sim1\degr$ west and north of NGC~2442, and
including NGC~2442 itself. These observations quite clearly show the
\ion{H}{1} in NGC~2442, but fail to detect any sign of HIPASS
J0731--69, since the $3\sigma$ column density achieved with the
limited number of baselines and amount of time available (i.e.,
$\sim10^{20}$~cm$^{-2}$) is still an order of magnitude greater than
the peak column density found in the HIPASS data. We infer from this
that the \ion{H}{1} seen by HIPASS is indeed somewhat diffuse, and not
clumpy (on the arcminute scales to which this ATCA configuration is
most sensitive) as one might perhaps expect if it was associated with
any particular optically-visible galaxy.

There are no cataloged galaxies within a $15\arcmin$ radius of the
central position of HIPASS J0731--69 in the {\em NASA Extragalactic
Database\/} (NED). We have examined the {\em Digitised Sky Survey\/}
for potential optical counterparts to the newly-discovered
gas. Although a total of seven independent objects were found within
$15\arcmin$ of the central position, none are more than $1\arcmin$ in
diameter, making it rather unlikely that any are new members of the
NGC~2442 group \citep{lgg93}. The most likely candidate object, an
isolated spiral galaxy at $\alpha=07^{\rm h} 31^{\rm m} 10\fs4$,
$\delta= -68\degr 55\arcmin 59\arcsec$ (J2000), has been found to have
a redshift inconsistent with that of the NGC~2442 group (I.~Perez,
private communication).

A channel map sequence (Figure~\ref{f:chanmaps}) shows that even with
the large HIPASS beamwidth and coarse velocity resolution, HIPASS
J0731--69 is resolved into multiple components. Two components,
centered on $(\alpha, \delta) = (07^{\rm h} 34^{\rm m}, -69\degr
04\arcmin)$ and $(07^{\rm h} 31^{\rm m}, -69\degr 20\arcmin)$ (J2000),
begin to appear at $V_{\odot} \sim 1325$~\kms\ and then grow
separately, before merging at a velocity near 1440~\kms.  A Gaussian
fit to this merged component yields a centroid of $\alpha=07^{\rm h}
31^{\rm m} 39\fs4$, $\delta= -69\degr 01\arcmin 36\arcsec$ (J2000). At
velocities near 1340~\kms and 1380~\kms, there are the first signs of
a direct connection of the emission from the northeast and the
southwest components respectively, to that from NGC~2442, and by
1418~\kms, both components are apparently joined with NGC~2442 as well
as with each other. The merged component of HIPASS J0731--69 again
appears well connected with NGC~2442 for most velocities in the range
1500--1600~\kms. However, it must be borne in mind that the separation
of the two main components, as well as their distance from NGC~2442,
are only $1-2\times$ the HIPASS beamwidth, so any apparent physical
connections between each of them must be regarded with caution until
higher resolution observations can be made.

\newpage
\section{Discussion}

Since the bulk of this new gas lies at a projected separation of
nearly $40\arcmin$ from NGC~2442 (and only slightly further from
NGC~2397), it is probably best characterized as an intergalactic cloud
associated with the NGC~2442 group, though both its velocity range,
and Fig.~\ref{f:h1ondss}, suggest the closest connection is to
NGC~2442. One way to set limits on the actual separation of HIPASS
J0731--69 from NGC~2442 would be to search for \Halpha\ emission
from the cloud's surface, due to ultraviolet photons escaping from
the galaxy. \Halpha\ emission levels of 2--4~milliRayleighs are 
expected for a cloud within 200~kpc of the galaxy \citep{jbh98}, and
emission measures as low as 1~milliRayleigh are now detectable
with 4~m-class telescopes (J.~Bland-Hawthorn, private communication).

Similar examples in nearby galaxy groups of intergalactic gas clouds
without prominent optical counterparts include: the ``Virgo Cloud''
\ion{H}{1} 1225+01 \citep{gh89}; the ``Leo Ring'' \citep{sch83}; the
\ion{H}{1} cloud in the NGC~3256 Group \citep{eng94}; and the
\ion{H}{1} cloud superimposed on the blue compact dwarf FCC~35 in the
Fornax Cluster \citep{mep98}.  Table~\ref{t:clouds} compares the
global properties of these four systems with HIPASS J0731--69.  In
each of the first two cases, faint dwarf galaxies apparently
associated with these gas clouds have since been identified
\citep{cgh95,sch89}. In the absence of either a contrast-enhanced
photo or a deep CCD image of the area covered by HIPASS J0731--69, we
cannot rule out the possibility of a ``Malin 1''-like object
\citep{ma1}, with an extended low surface brightness disk and very
compact bulge component. However, the failure to detect gas clumping
with the ATCA makes it rather less likely that any optical (galactic)
counterpart exists.

The complex velocity field of HIPASS J0731--69 illustrated by
Fig.~\ref{f:chanmaps}, and the rather asymmetric \ion{H}{1} profile of
Fig.~\ref{f:h1prof}, are inconsistent with a uniformly rotating disk
structure. Rather, it is more likely that HIPASS J0731--69 is an
extended, stream-like feature of some kind. The \ion{H}{1}~velocity
field for NGC~2442 of \cite{sjh98} shows line-of-sight velocities
ranging from just below 1100~\kms\ in the northeastern quadrant, to
over 1700~\kms\ in the southwest. However, gas in the northern arm
with velocities in the approximate range of $1450-1600$~\kms\ appears
to be being drawn out and away from the rest of the disk in an
easterly direction, toward the southwest component of HIPASS
J0731--69. In addition, sections of the inner northern arm are
displaced in velocity from the background disk by $\sim50$~\kms.  Both
these distortions are consistent with gas being pulled out of the
plane, and their respective velocities and alignments lead us to
hypothesize that they may, with better sensitivity and resolution,
prove to be contiguous with the various components of HIPASS
J0731--69. Similar anomalous velocities and double line profiles in
the northern arm were seen optically by \citet{bap99} and by
\citet{mb97}.

HIPASS J0731--69 spans $\sim$180~kpc, and contains
$10^{9}$~M$_{\odot}$ of atomic hydrogen, making it comparable in scale
to both the Virgo Cloud and the Leo Ring. The Leo Ring contains as
much gas as either of the nearby spiral galaxies M95 or M96, while the
Magellanic system (LMC, SMC, Stream, Bridge, and Leading Arm) contains
$\sim10^{9}$~M$_{\odot}$ of \ion{H}{1} \citep{mep01}, or one-third as
much gas as the Milky Way. The ``forked'' velocity structure of HIPASS
J0731--69 is also not unlike that of the Magellanic
Stream/Bridge/Leading Arm system \citep{put98}, or the sections of the
Leo Ring studied by \citet{sch85}, which he was able to model quite
well by a Keplerian orbit. The spatial resolution of the HIPASS data
is insufficient to allow us to test any orbit models, but future
mapping at higher resolution may enable the internal dynamics of this
system to be understood.

Was the gas in HIPASS J0731--69 once part of NGC~2442? The \ion{H}{1}
mass-to-blue luminosity ratio for NGC~2442 alone ($\log \langle {\rm
M}_{\rm HI} / {\rm L}_{B} \rangle = -0.57$, using our \ion{H}{1} mass
and $B_{\rm T}=11.1$ \citep{sd93}), is identical to the median value
for Sbc galaxies as a whole \citep{rh94}. However, compared with other
Sbc galaxies, NGC~2442 is slightly sub-luminous optically, and its
\ion{H}{1} content is also just below the median. Including HIPASS
J0731--69 would then make the total atomic gas content of NGC~2442
about average for an Sbc galaxy (this of course assumes that the
disturbance to NGC~2442 has not radically changed its apparent Hubble
type). The scatter in each parameter exhibited by Sbc galaxies
($\sim0.5$~dex) means that global parameters are not particularly
helpful in settling the question of the origin of HIPASS J0731--69.

We now briefly review how our discovery of HIPASS J0731--69 impacts
upon the two main theories for the origin of the asymmetry in
NGC~2442.

\subsection{Tidal encounter?}

As mentioned in the Introduction, there are several candidate galaxies
for a recent past interaction with NGC~2442. What kind of close
encounter might have caused NGC~2442 to lose up to 25\% of its own
\ion{H}{1} content (or strip an equivalent amount of gas from some
other galaxy), while only mildly affecting its global kinematics, infrared
luminosity, and star formation rate \citep{mb97}?

Large tidal tails and bridges on scales of 100~kpc or more are not
uncommon in compact groups (e.g., the Leo triplet; Haynes, Giovanelli,
\& Roberts 1979), and in galactic mergers (see e.g., the merger
``sequence'' imaged in \ion{H}{1} by \citet{hvg}). In such cases
however, there are usually two or more galaxies of comparable mass
involved, with separations of this same order. In addition, there is
growing evidence for the formation of ``tidal dwarf galaxies'' from
the gas in these tails \citep{dm99}. The resolution of the HIPASS data
is not really adequate for us to determine whether HIPASS J0731--69 is
an isolated cloud (or group of clouds), or if there is a bridge of
material linking it directly to NGC~2442. We note that \citet{sjh98},
in her ATCA study of NGC~2442, observed a mild ``stretching'' of the
\ion{H}{1} disk towards the west, but did not find any signs of an
\ion{H}{1}~tail within the $34\arcmin$ primary beam, making the former
more likely.

\citet{mb97} provide a model for a close passage of AM 0738--692
$\sim$200~Myr ago, which appears to be reasonably successful at
explaining many of the morphological peculiarities of NGC~2442. There
is no specific prediction of a large gas cloud like HIPASS J0731--69
in this scenario (but then neither was there any particular reason to
expect such a feature). According to this model, the northern arm in
particular is not driven by a spiral density wave, but is more like a
tidal tail. Another good example of this phenomenon is the interacting
pair NGC~6872/IC~4970 \citep{mbr93} which, just as in NGC~2442,
exhibits large velocity gradients across the width of the tidal arm as
material originally drawn away from the disk begins to fall back in,
is compressed and shocked.

This mechanism accounts quite naturally for the prominent dust lane,
enhanced star formation, and unusual line profiles observed in the
northern arm of NGC~2442. However, in the \citet{mb97} model, it is
hard to believe that such an innocuous object as AM 0738--692 could
once have contained as much gas as is in HIPASS J0731--69, or that it
could have removed such a large quantity of gas to such a large
distance. If NGC~2442 has been the victim of a recent tidal encounter,
then the existence of HIPASS J0731--69 makes it far more likely that
the culprit is a more massive galaxy to the northwest, such as the
elliptical galaxy NGC~2434 or perhaps even the spiral + Magellanic
Irregular pairing of NGC~2397 and NGC~2397A.


\subsection{Ram pressure stripping?}

The NGC~2442 system shares many of the characteristics of certain
asymmetric galaxies, which are suspected of being ram-pressure
stripped of some of their \ion{H}{1} as they pass through the
intracluster medium. Among the best examples of this phenomenon are
NGC~2276 \citep{gru93}, NGC~4273 \citep{dav95}, NGC~7421
\citep{sdr97}, and the Virgo cluster galaxies NGC~4388 \citep{vbh99},
NGC~4654 \citep{pm95}, and NGC~4522 \citep{vol01}. The most
distinctive features of these objects are: (a) strongly asymmetric
\ion{H}{1} profiles; (b) a truncated optical and/or gas distribution
on one side of the galaxy, often having a ``bow shock''-like
morphology; (c) a gaseous tail or ``wake'' in the opposite direction,
possibly out of the disk plane; and (d) mild disturbance to an
otherwise regular rotational velocity pattern. In some cases, the
presence of a diffuse, hot gas component is confirmed by X-ray
satellite observations, but the intracluster medium is not always so
obvious.

The deep optical image of NGC~2442 (Fig.~\ref{f:dfm}) shows quite a
sharp edge to the northern edge of the stellar disk, when compared
with the diffuse extensions to the southeast and southwest. The global
asymmetry and sharp cutoff to the north side is even more pronounced
in H$\alpha$ \citep{rd93,mb97}.  The northern arm may represent the
bow shock, where the disk is ploughing into the densest part of the
inter-group medium, resulting in the observed velocity jumps. Disk
rotation would stretch out the shock front, with the leading edge at
the eastern end of the northern arm, and material flowing downstream
along this arm to the west and south (this assumes the southern side
of the disk of NGC~2442 is in the foreground). HIPASS J0731--69 would
then represent a stream of gas trailing out behind (or completely
detached from) NGC~2442, perhaps from a time when the galaxy
experienced a particularly strong ram pressure. Both NGC~2442, and one
of its companions to the south (ESO 059-G006), show distortions in
their \ion{H}{1} velocity fields \citep{sjh98} which could just as
easily be due to their passage through some inter-group medium, as to
an interaction. A {\em ROSAT} HRI image barely detects NGC~2442 itself
however, let alone any hot, diffuse gas.

It is interesting to contrast the global \ion{H}{1} profile of
NGC~2442 (Fig.~\ref{f:h1prof}) with the global $^{12}$CO ($1-0$)
profile of \citet{bco95}; the asymmetry in the atomic gas distribution
is {\em reversed\/} in the molecular gas distribution, with the bulk
of the CO emission apparently coming from the northeast sector of the
disk, at lower velocities. \citet{bco95} did not survey the entire
disk of NGC~2442, but were unable to detect $^{12}$CO ($1-0$) much
beyond the point where the northern arm bends back on itself. Since
the molecular gas is predominantly located in the inner disk, where
the total mass surface density is highest (and ISM stripping least
effective), any asymmetry in the CO distribution would tend to imply
the action of tidal forces, rather than ran-pressure
\citep{com88,bos94}.  Since the ratio of molecular-to-atomic gas in
NGC~2442 ($M_{\rm H2}/ M_{\rm HI} \sim 0.7$) is close to the average
for Sbc galaxies \citep{ys91}, this too weakens slightly the case for
ram pressure stripping.

Clearly, HIPASS J0731--69 is an important new clue in the puzzle of
NGC~2442's disturbed appearance, as well as being a significant object
in its own right on account of its apparent rarity in the HIPASS
survey. Followup observations with more compact configurations of the
ATCA are planned, as well as higher spectral resolution observations
with the Parkes multi-beam system, and will be crucial in telling us
more about the origin of HIPASS J0731--69 and its relationship to
NGC~2442.

\acknowledgments

The Parkes telescope and Australia Telescope Compact Array are both
part of the Australia Telescope, which is funded by the Commonwealth
of Australia for operation as a national facility managed by CSIRO.
We are grateful to the staff at the ATNF Parkes and Narrabri
observatories, and to members of the Zone of Avoidance survey team,
for assistance with HIPASS and follow-up observations. We especially
wish to thank I.~Perez, N.~Killeen, T.~Getts, and
J.~Bland-Hawthorn. This research has made use of the NASA/IPAC
Extragalactic Database (NED) which is operated by the Jet Propulsion
Laboratory, California Institute of Technology, under contract with
the National Aeronautics and Space Administration.  The Digitized Sky
Surveys were produced at the Space Telescope Science Institute under
U.S. Government grant NAG W-2166. The images of these surveys are
based on photographic data obtained using the Oschin Schmidt Telescope
on Palomar Mountain and the UK Schmidt Telescope. The UK Schmidt
Telescope was operated by the Royal Observatory Edinburgh, with
funding from the UK Science and Engineering Research Council (later
the UK Particle Physics and Astronomy Research Council), until 1988
June, and thereafter by the Anglo-Australian Observatory.




\vspace{3cm}

\figcaption[h1onb.eps]{\protect{Integrated \ion{H}{1}} column density
contours from the HIPASS observations of the region around NGC~2442,
overlaid on a $B$-band image
from the {\em Digitized Sky Survey}, together with grid lines of
constant Right Ascension and Declination. The contours are
plotted at levels of 0.01, 0.02, 0.03, 0.06, 0.09, 0.2, and
0.4~M$_{\odot}$~pc$^{-2}$ (multiply by $1.25\times10^{20}$ to get
equivalent number of H atoms per cm$^{2}$).
Note the disturbed appearance of the spiral arms in NGC 2442, the
absence of any significant \ion{H}{1} contribution from NGC~2434 (just
left of center), and the lack of a bright optical counterpart to
HIPASS J0731--69. \label{f:h1ondss}}

\vspace{3cm}

\figcaption[dfm2442lab.eps]{Deep optical image of the field immediately
surrounding NGC~2442,
obtained by summing three blue photographic plates from the UK Schmidt
Telescope in the manner outlined by \citet{dfm81}. Comparison with the
surface photometry of \citet{sd93} indicates that the faintest structures
visible have a surface brightness of $B \sim 28$~mag~arcsec$^{-2}$.
With the exception of NGC~2442 at the center, all galaxies listed in
Table~\protect{\ref{t:neigh}} which are visible in this image are
identified by a label immediately above each one. The solid bar at
lower right is $5\arcmin$ in length.
\label{f:dfm}}

\clearpage

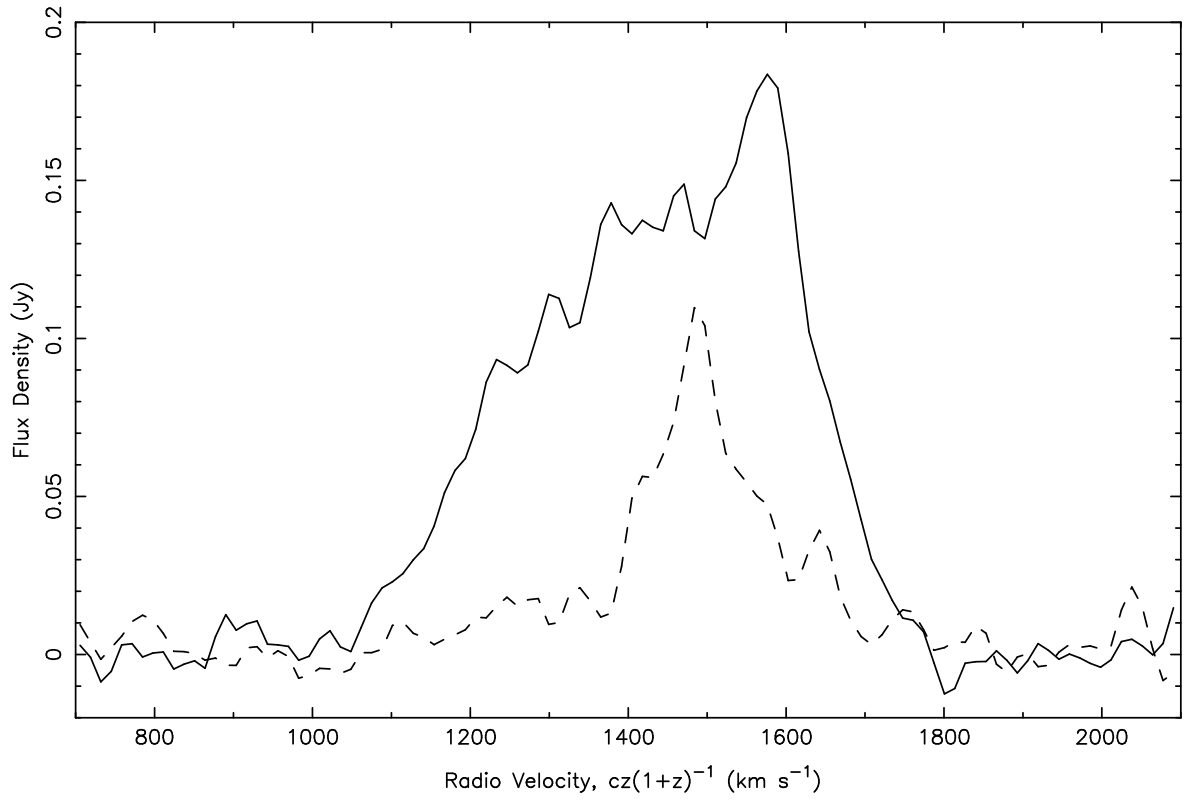
\figcaption[newprof.ps]{\protect{\ion{H}{1}} line profiles of NGC~2442
({\em solid
line}), and HIPASS J0731--69 ({\em dashed line}) from the HIPASS data.
In each channel, the emission in a $28\arcmin \times 28\arcmin$ box has
been summed. The line profiles have been continuum-subtracted, and then
Hanning-smoothed.
\label{f:h1prof}}
\includegraphics{f3.eps}

\clearpage

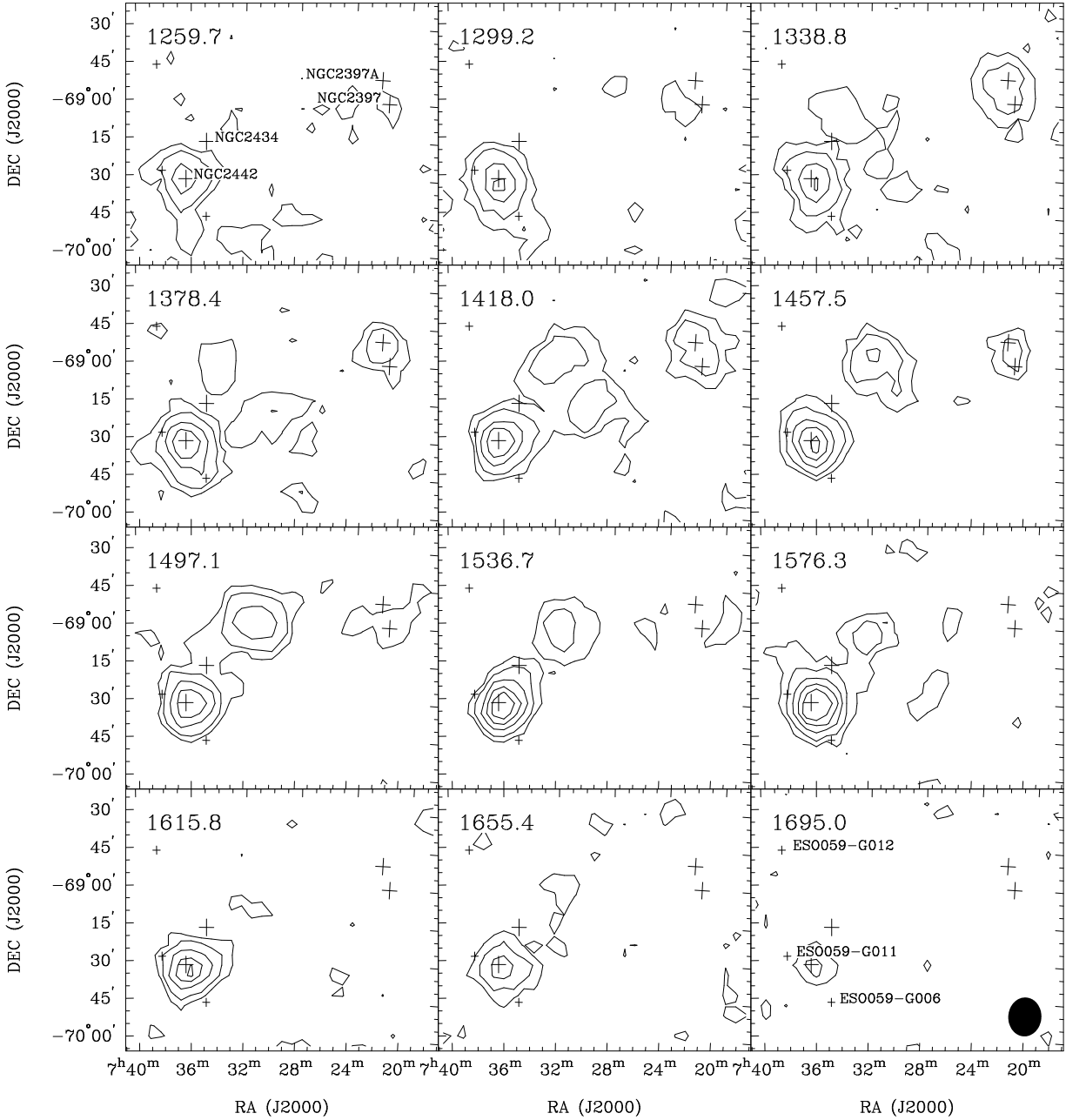
\figcaption[chanmaps.ps]{Channel maps of \ion{H}{1} emission from
the HIPASS observations of the region around NGC~2442.
The contours correspond to flux densities of 15, 30, 60, 90, and
120~mJy~beam$^{-1}$. For reference, the optical positions of the
four major NGC galaxies in the field are marked in each panel with
a large cross, and identified in the first panel; three more ESO
galaxies are marked with small crosses, and identified in the last
panel. The HIPASS beamsize is also indicated in the last panel.
\label{f:chanmaps}}
\includegraphics{f4.eps}





\clearpage

\begin{deluxetable}{lcccl}
\tabletypesize{\scriptsize}
\tablecaption{Neighbors of NGC 2442 with known redshift.
\label{t:neigh}}
\tablewidth{0pt}
\tablehead{
\colhead{Galaxy} & \colhead{$\alpha$ (J2000)} & \colhead{$\delta$ (J2000)} &
\colhead{$V_{\odot}$}   & \colhead{Source} \\
\colhead{~~}  & \colhead{~~}  &  \colhead{~~} & \colhead{(\kms)}           &
\colhead{~~}
}
\startdata
NGC~2442 &
$07^{\rm h} 36^{\rm m} 23\farcs9$ & $-69\degr 31\arcmin 48\arcsec$ &
1475 & \citet{mb97} \\
LEDA 100030 &
$07^{\rm h} 36^{\rm m} 40\farcs7$ & $-69\degr 26\arcmin 39\arcsec$ &
1331 & \citet{sjh98} \\
AM 0738--692 &
$07^{\rm h} 38^{\rm m} 11\farcs8$ & $-69\degr 28\arcmin 27\arcsec$ &
1529 & \citet{mb97} \\
AM 0737--691 & 
$07^{\rm h} 37^{\rm m} 12\farcs6$ & $-69\degr 20\arcmin 31\arcsec$ &
1456 & \citet{sjh98} \\
NGC 2434 &
$07^{\rm h} 34^{\rm m} 51\farcs4$ & $-69\degr 17\arcmin 01\arcsec$ &
1390 & \citet{rc3} \\
ESO 059-G006 &
$07^{\rm h} 34^{\rm m} 51\farcs1$ & $-69\degr 46\arcmin 49\arcsec$ &
1346 & \citet{sjh98} \\
ESO 059-G012 &
$07^{\rm h} 38^{\rm m} 31\farcs8$ & $-68\degr 46\arcmin 16\arcsec$ &
1323 & \citet{hdn96} \\
NGC 2397 &
$07^{\rm h} 21^{\rm m} 20\farcs8$ & $-69\degr 00\arcmin 07\arcsec$ &
1363 & \citet{mfb92} \\
NGC 2397A &
$07^{\rm h} 21^{\rm m} 56\farcs3$ & $-68\degr 50\arcmin 45\arcsec$ &
1392 & \citet{bgc01} \\
\enddata
\end{deluxetable}

\begin{deluxetable}{lrrrcrl}
\tabletypesize{\scriptsize}
\tablecaption{Global Properties of Some Extragalactic \ion{H}{1} Clouds.
\label{t:clouds}}
\tablewidth{0pt}
\tablehead{
\colhead{Source} & \colhead{$V_{\odot}$}   & \colhead{$\int S_{\rm HI} dV$} &
\colhead{$D$}   & \colhead{$M_{\rm HI}$}  & \colhead{Extent} & \colhead{Ref} \\
\colhead{~~}     & \colhead{(\kms)}          &  \colhead{(Jy~\kms)}           &
\colhead{(Mpc)} & \colhead{($10^{8}$~M$_{\odot}$)} & \colhead{(kpc)} &
\colhead{~~}
}
\startdata
HI 1225+01          & 1275  & 42.4 & 20.0 & 40   & 200 & \citet{gh89} \\
Leo Ring            &  700  & 70.9 & 10.0 & 17   & 260 & \citet{sch89} \\
NGC 3256 Cloud      & 2868  & 15.0 & 37.0 & 48   & 110 & \citet{eng94} \\
FCC 35 Cloud        & 1658  &  2.9 & 18.2 & 2.2   &  13 & \citet{mep98} \\
HIPASS J0731--69    & 1481  & 18.0 & 15.5 & 10   & 180 & This work \\
\enddata
\end{deluxetable}



\end{document}